%% file: main.tex
\documentclass[10pt,letterpaper,compsoc,conference]{iiswc23}

\usepackage{cite}
\usepackage{amsmath,amssymb,amsfonts}
\usepackage{algorithmic}
\usepackage{graphicx}
\usepackage[dvipsnames]{xcolor}
\usepackage[final]{microtype}
\usepackage[italic]{mathastext}
\usepackage{libertine}
\usepackage[T1]{fontenc}
\usepackage{textcomp}
\usepackage[varqu,varl]{zi4}
\usepackage[all]{nowidow}
\usepackage[keeplastbox]{flushend}

\usepackage{svg}

\usepackage{dblfloatfix}    
\usepackage{enumitem}
\usepackage{framed}
\usepackage{multirow}
\usepackage{url}
\usepackage{balance}
\usepackage{booktabs}
\usepackage{listings}
\usepackage{caption}
\usepackage{upquote}
\usepackage{authblk}
\usepackage{xcolor}
\usepackage{comment}
\usepackage{subfigure}
\usepackage{mathtools}
\usepackage{ragged2e}
\usepackage{tikz}
\usepackage{graphicx}
\usepackage{listings}
\usepackage{xcolor}
\usepackage{mathtools}
\usepackage{booktabs}
\usepackage{listings,newtxtt}
\usepackage[colorlinks = true,
        linkcolor = blue,
        urlcolor  = blue,
        citecolor = red,
        anchorcolor = blue]{hyperref}
\usepackage{xspace}

\lstdefinestyle{mystyle}{
    backgroundcolor=\color{backcolour},
    commentstyle=\color{codegreen},
    keywordstyle=\color{magenta},
    numberstyle=\tiny\color{codegray},
    stringstyle=\color{codepurple},
    basicstyle=\ttfamily\footnotesize,
    breakatwhitespace=false,
    breaklines=true,
    captionpos=b,
    keepspaces=true,
    numbers=left,
    numbersep=5pt,
    showspaces=false,
    showstringspaces=false,
    showtabs=false,
    tabsize=2
}
\newcommand{\papertitle}{CXL-ClusterSim\xspace}

\definecolor{codegreen}{rgb}{0,0.6,0}
\definecolor{codegray}{rgb}{0.5,0.5,0.5}
\definecolor{codepurple}{rgb}{0.58,0,0.82}
\definecolor{backcolour}{rgb}{0.95,0.95,0.92}

\definecolor{lightgray}{gray}{0.8}
\newcommand\encircle[1]{%
	\tikz[baseline=(X.base)] 
	\node (X) [draw, shape=circle, scale=0.7, inner sep=0, fill=black, text=white] {\strut #1};%
}

\newcommand\wcircle[1]{%
	\tikz[baseline=(X.base)] 
	\node (X) [draw, shape=circle, scale=0.7, inner sep=0, fill=white, text=black] {\strut #1};%
}
\AtBeginDocument{%
  \providecommand\BibTeX{{%
    \normalfont B\kern-0.5em{\scshape i\kern-0.25em b}\kern-0.8em\TeX}}}

\begin{document}


\title{\papertitle: Modeling CXL-based Disaggregated Memory Cluster for Pooling and Sharing using gem5 and SST}





\author{Kaustav Goswami}
\author{Maryam Babaie}
\author{Hoa Nguyen}
\author{Venkatesh Akella}
\author{Jason Lowe-Power}
\affil{University of California, Davis}

\maketitle
\pagestyle{plain}


\input{Text/abstract}
\input{Text/intro_v2}

\input{Text/background_v2}
\input{Text/method_v3}
\input{Text/eval_v2}

\input{Text/related}
\input{Text/conclusion}
\input{Text/ack}

\bibliographystyle{IEEEtranS}
\bibliography{reference}


\end{document}

%% file: Text/abstract.tex
\begin{abstract}
    Large-scale AI training and inference require hundreds of gigabytes to terabytes of DRAM with high peak to average utilization ratios, resulting in overprovisioning.
    In cloud computing, DRAM constitutes a significant share of the cost.
    Yet, as shown by recent articles, DRAM is heavily under utilized.
    Memory disaggregation is a solution to both these problems.
    With the advent of the CXL protocol, there is renewed interest in designing and optimizing computing systems with disaggregated memory.
    However, at present, there are limited simulation tools available for exploring the design space and evaluating the performance tradeoffs in computer systems with disaggregated memory.
    
    In this paper, we propose \papertitle, a full-system modeling and simulation framework by combining the gem5 simulator for fidelity, with the Structural Simulation Toolkit (SST) for parallel simulation. 
    We outline the challenges in creating this simulation infrastructure and present a design that is scalable, flexible, and reasonably fast to help computer architects to explore the design space of CXL-based disaggregated memory and identify new opportunities for hardware/software codesign and performance optimization.
\end{abstract}

%% file: Text/intro_v2.tex
\section{Introduction}
\label{sec:intro}
Emerging applications in scientific computing and artificial intelligence applications are memory-intensive.
For instance, GPT-4 has 1.76 trillion parameters, and the number of parameters in large transformer models is expected to grow by a factor of 410$\times$~every two years~\cite{gpt4-params,gholami2024ai}.
However, memory performance and capacity have increased at a modest pace due to the challenges in memory scaling~\cite{memory-scaling}.
To make matters worse, the ratio of the peak to average memory demand in training large ML models is high.
That means systems are typically overprovisioned~\cite{pond}, which adds to the cost and power consumption.
For example, Microsoft reports up to 25\% of the DRAM capacity becomes stranded in their Azure cloud servers~\cite{pond}.

Researchers propose memory disaggregation, \textit{i.e.}, the decoupling of compute and memory resources, as a solution to these challenges~\cite{mem-disaggregation,lim2012system}.
In this architecture, memory is typically \textit{pooled} out of a memory blade, where physical capacity is aggregated into a central reservoir for dynamic allocation to specific compute nodes~\cite{starnuma}, or \textit{shared}, allowing multiple nodes to concurrently access the same address space for low-latency data exchange~\cite{cxl-3.1-spec}.
By enabling both precise resource partitioning and multi-node collaboration, this approach addresses modern workload requirements while significantly mitigating memory underutilization.


Though the idea of memory disaggregation has been around for many years~\cite{lim2012system}, the advent of Compute Express Link (CXL) protocol has made the design and implementation of disaggregated memory systems practical.
As an open standard built upon the PCI Express (PCIe) infrastructure, CXL facilitates high-speed, coherent communication between CPUs and memory devices.
With its low latency and high bandwidth capabilities, CXL allows for the creation of flexible memory pools that can be dynamically allocated to different processors in a network, effectively addressing the challenges of memory stranding and underutilization in data centers and cloud services.

Despite growing research into CXL-based disaggregated memory, current modeling infrastructures often lack the full-system fidelity required to accurately capture the performance nuances of these emerging architectures.
Gouk \textit{et al.} used FPGA-based prototypes, which are based on older CXL specifications, hard to access and modify, and not scalable to multiple nodes~\cite{gouk2022direct}.
Puri \textit{et al.} introduced DRackSim~\cite{dracksim}, a simulation framework built on gem5~\cite{gem5} and DRAMSim2~\cite{dramsim2} to model memory disaggregation.
However, the framework relies on gem5's Syscall Emulation (SE) mode.
Therefore, it cannot account for operating system (OS) overheads or complex kernel-level memory management behaviors.

In another direction, researchers have used NUMA nodes to emulate disaggregated memory~\cite{pond, tpp}.
However, studies show that using NUMA nodes as a proxy for CXL-based disaggregated memory is not ideal and reveal some differences between real CXL devices and NUMA-emulated CXL~\cite{demystifying-paper}.


In this work, we propose \textit{\papertitle}, a scalable and flexible simulation infrastructure to model and assess the performance of CXL-based disaggregated memory systems through full-system simulation.
Simulation is an indispensable tool in the architectural development of future systems due to its accessibility, scalability, and ability to model complex systems cost-effectively.
The current research landscape lacks comprehensive frameworks for the hardware/software co-design of CXL-based disaggregated memory, hindering the design-space exploration.
Moreover, CXL specifications are written with functionality in mind, and there is a need to evaluate the practical performance of these theories.
\papertitle~aims to fill this gap by providing a platform to evaluate the performance of CXL-based disaggregated memory systems.

\papertitle~is built on top of two popular open-source simulators, gem5~\cite{gem5, gem5-memctrl} and the Structural Simulation Toolkit (SST)~\cite{sst}.
gem5 is used to model full-system compute nodes (\textit{i.e.}, \textit{hosts} in CXL specification, \textit{system node} in this work) with detailed timing-models, providing a comprehensive view of system behavior, including OS impact, TLBs, huge pages, etc.
Each system node consists of multiple cores, a cache hierarchy, and a memory system, running a separate OS and application.
SST, on the other hand, is primarily used to simultaneously simulate multiple gem5 nodes, making the simulation of a cluster of system nodes feasible in real time.
In addition, SST also models the disaggregated memory blade, referred to as the \textit{remote memory node}.
In this work, we demonstrate the capabilities of \papertitle, \textit{i.e.}, pooling and sharing, by running a set of full-system experiments.
\begin{itemize}
  \item First, we run the STREAM benchmark~\cite{stream-article,stream-techrep} to evaluate the memory bandwidth of the system by varying different NUMA policies.
  Alongside validating our simulation model, we also present results showing whether system-level effects are captured.
  We study the scalability of the entrire cluster by varying the number of hosts.
  In addition, we show memory pooling with heterogeneous ISAs at two different hosts (ARM and RISCV). 
  \item Second, we demonstrate the ability to run real workloads by running a multiprogrammed example with the NAS Parallel Benchmarks (NPB)~\cite{npb}.
  We present a case study of memory stranding effects when NUMA \textit{preferred} is used for the local memory, while the majority of the data is pooled from a 128 GiB device.
  \item Third, we show memory sharing using GAPBS~\cite{gapbs}.
  We modify the allocation of the graphs to use the remote memory.
  Several hosts share a graph stored on the remote memory node and execute a graph kernel on the system node.
  Here we present a \textit{single writer multiple reader} sharing model.
\end{itemize}

The main contribution of the paper is a full-system simulation and modeling framework for CXL-based disaggregated memory.
The framework is a combination of current \textit{state-of-the-art} simulators that will enable future hardware-software co-design research for the computer architecture and computer systems communities.
\papertitle~is not tied to PCIe or link-level details in the CXL specification.
It is flexible to explore different implementations of disaggregated memory~\cite{opencapi,genz,cxl-3.1-spec}, including the CXL specification and changes to the future CXL specifications.
This tool will be released as open-source, and we will work with the gem5 and SST communities to have our extensions merged with the mainline code to the community to enable further research in the area of CXL-based disaggregated memory systems.
There are still many more features that can be added to the framework, such as performance/parallel efficiency, CXL table structures, \textit{etc.}, and we look forward to contributions from the community to improve the framework.

This paper is organized as follows.
Section~\ref{sec:background} describes the background of memory disaggregation and the current simulation infrastructure.
We explain how \papertitle~works in Section~\ref{sec:method}, followed by validation and evaluation in Section~\ref{sec:eval}.
Section~\ref{sec:related} briefly explains the related works towards simulating memory disaggregation.
We conclude the paper in Section~\ref{sec:conclusion} with further scopes of improvements.


%% file: Text/background_v2.tex
\section{Background and Motivation}
    \label{sec:background}
    In this section, we provide a background of each of the two simulators that we use and explain the basics of the CXL protocol.

    \subsection{Compute Express Link (CXL)}
        \label{sec:background-cxl}
        \subsubsection{CXL Overview and Protocols}
            Compute Express Link (CXL) is an open-standard interconnect designed to facilitate high-speed, low-latency communication between processors, accelerators, and memory expansion devices~\cite{dassharmax,cxl-main,cxl-3.1-spec,cxl-4-spec}.
            It leverages the PCIe physical layer.
            With CXL 3.1 aligning with PCIe 6.0 to provide up to 64 GT/s per lane~\cite{cxl-3.1-spec}, CXL maintains memory coherency between the host and attached devices~\cite{rambus-cxl}.
            The protocol stack comprises three sub-protocols:
            (1) \texttt{CXL.io} for device discovery and configuration;
            (2) \texttt{CXL.cache} for device-side caching of host memory;
            (3) \texttt{CXL.mem}, which enables the host to access device memory using standard load/store instructions.

        \subsubsection{Definitions and Architecture}
            The CXL specification defines \textit{hosts} and \textit{devices}.
            Hosts are compute nodes with or without local memory.
            The devices, on the other hand, are remote memory nodes.
            There is a global address space where devices are mapped.
            Hosts may choose to map a part of their own address space into the global address space.
            The global address space is one contiguous address space.
            The fabric manager, which is the application-specific logic responsible for orchestrating and managing the control plane of a CXL fabric.
            In addition, it is tasked with binding hosts and devices to the global address space.

            \texttt{CXL.mem} supports two distinct memory utilization modes:
            \begin{itemize}
                \item \textit{Memory Pooling}: The device memory is partitioned into distinct segments, each exclusively assigned to at most a single host.
                This mitigates stranded memory by allowing the fabric manager to reassign capacity based on demand.
                \item \textit{Memory Sharing}: Introduced in the CXL 3.0 specification~\cite{cxl-3.1-spec}, this allows multiple hosts to concurrently access the same physical memory range.
                The protocol provides Back-Invalidate snoops at the cache-line granularity.
            \end{itemize}

            Table~\ref{tab:abbreviation} lists all the terms used for this paper and their correspondence to the CXL specification.

            \begin{table}[t]
                \centering
                \resizebox{0.47\textwidth}{!}{%
                \begin{tabular}{ccc}
                    \hline
                    \textbf{Term} & \textbf{Meaning} & \textbf{CXL Spec}\\ \hline \hline
                    system node & \begin{tabular}[c]{@{}c@{}}A fully-fledged compute node\\ with independent cores, caches, \\ and, operating systems.\end{tabular} & host \\ \hline
                    remote memory node & A memory blade. & device memory \\ \hline
                    local memory & \begin{tabular}[c]{@{}c@{}}The main-memory attached\\ with the system node.\end{tabular} & local memory \\ \hline
                    remote memory & Memory inside a memory blade. & \begin{tabular}[c]{@{}l@{}}host-managed\\ device memory\end{tabular} \\ \hline
                    \end{tabular}%
                }
                \caption{Terminologies used in this paper.
                We define and map them to the CXL specifications.
                These terms are used interchangeably in this paper.}
                \label{tab:abbreviation}
            \end{table}

        \subsubsection{Scope of this Work}
            \label{sec:scope}
            The framework provided in this paper is designed to evaluate the \textit{performance} of \texttt{CXL.mem}-based disaggregated memory systems only.
            Although \texttt{CXL.io} is used to configure the interface between the host and the device, we keep \texttt{CXL.io} as future work since the initialization phase has a negligible impact on the overall performance.

    \subsection{Simulators}
        \subsubsection{gem5 Simulator}
            \label{sec:background-gem5}
            The gem5 simulator is an event-driven, advanced, modular platform for computer architecture research, capable of simulating a wide range of systems from small embedded devices to large-scale server architectures~\cite{gem5}.
            A \textit{SimObject} is the foundational C++ base class in gem5 representing a hardware component (such as a CPU, cache, or memory controller) that can be instantiated, configured via Python, and managed by the simulation's event queue.
            Today, gem5 offers a variety of CPU models, ranging from simple in-order processors to complex out-of-order processors, enabling detailed studies of different microarchitectural features.
            These models support various instruction set architectures (ISAs) such as x86, ARM, RISC-V, \textit{etc.}, making gem5 a versatile tool for cross-ISA research and comparison.
            In addition to CPU models, gem5 supports detailed modeling of I/O devices like Graphics Processing Units (GPUs), essential for research in parallel processing, machine learning, and high-performance computing.
            
            gem5's memory subsystem provides high-fidelity models of memory technologies such as DDR, LPDDR, HBM, and non-volatile memory~\cite{gem5-memctrl}.
            The cache models in gem5 are highly configurable, allowing the simulation of different cache hierarchies, policies, coherence protocols like MESI, MOESI, and more, facilitating the exploration of various caching strategies.

            In this work, we leverage gem5's full-system mode for simulation.
            Full-system simulation enables the execution of complete software stacks, including OS and application software, on simulated hardware.
            For disaggregated memory research, we need to expose the OS interfaces for memory pooling, \textit{i.e.} NUMA and DAX respectively.
            This mode is particularly valuable for hardware-software co-design research, as it captures system-level interactions which allows researchers to evaluate how changes in hardware (or software) design affect software (or hardware) performance and behavior.
            By running real workloads and OS such as Linux, researchers can identify bottlenecks, optimize hardware-software interactions, and develop more efficient and robust systems.
            In this work, we boot an unmodified Ubuntu-based Linux distribution and run applications compiled on that system, creating an environment exactly the same as what will be found in a real system.


        \subsubsection{Structural Simulation Toolkit (SST)}
            \label{sec:background-sst}
            The Structural Simulation Toolkit (SST) is a comprehensive, highly scalable framework designed for the modeling and analysis of high-performance computing (HPC) systems~\cite{sst}.
            The architecture is divided into two primary layers: SST-Core, which provides the underlying event-driven simulation engine and synchronization API, and SST-Elements, a library of modular component models used to construct specific simulation targets.
            This modularity allows researchers to assemble interchangeable components, such as processors, memory hierarchies, and interconnects, into detailed, custom environments tailored to specific research needs.

            We use SST's Message Passing Interface (MPI) for communication between simulated nodes, enabling high-performance parallel execution across distributed computing clusters.
            Beyond standalone modeling, SST supports sophisticated co-simulation, allowing independent simulators to run concurrently and interact within a unified timeline.
            This capability is essential for capturing complex system-level interactions, such as how network topology impacts data movement or how memory hierarchy changes affect processor throughput.

            Furthermore, SST's extensibility allows it to integrate with external simulation tools to enhance its fidelity.
            For example, it can be bridged with gem5 to facilitate high-fidelity, full-system CPU-memory simulations~\cite{gem5sst}.
        
        \subsubsection{Addressing the Missing Gap}
            \label{sec:motivation}
            While gem5 provides high fidelity by allowing researchers to simulate an entire host in intricate detail, the overall simulation time is slow~\cite{gem5sst,mohammad2017dist,pd-gem5}.
            This is due to the single-threaded design of gem5~\cite{mohammad2017dist}.
            The simulation time of a cluster of gem5 hosts, running on a single thread, further exacerbates the  total simulation time when modeling disaggregated memory.
            On the other hand, SST provides scalability via the MPI infrastructure.
            However, it lacks the fidelity provided by a gem5 full-system simulation~\cite{gem5sst}.
            Therefore, by combining the two diverse tools, \textit{i.e.}, gem5 and SST, \papertitle provides a multi-scale environment capable of evaluating the intricate performance nuances of emerging architectures like CXL-based disaggregated memory.

%% file: Text/method_v3.tex
\section{Simulating Disaggregated Memory}
    \label{sec:method}

\begin{figure}[t]
        \centering
        \includegraphics[width=\columnwidth]{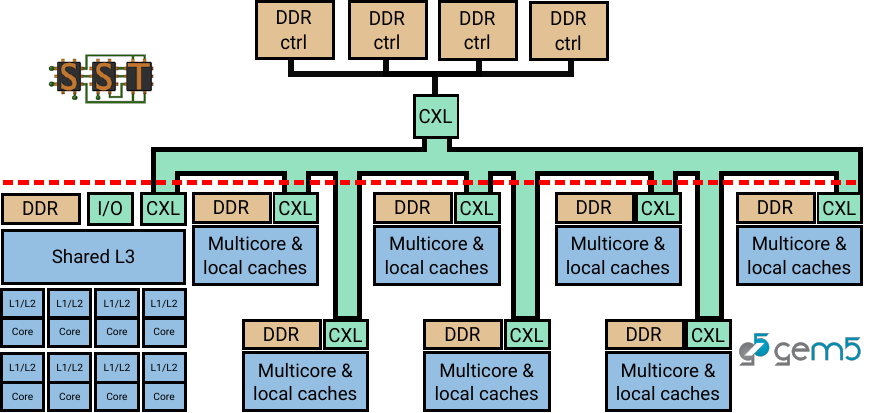}
        \caption{Overview of the simulation infrastructure.}
        \label{fig:gem5-sst}
    \end{figure}

    Figure~\ref{fig:gem5-sst} shows an overview of our new simulation infrastructure for modeling CXL-like disaggregated memory.
    We want to model multi-node CXL clusters with several hosts and devices.
    We use the gem5 simulator to model and simulate the hosts, and SST to model and simulate the CXL-like interconnect and the remote memory node, or the CXL device.

    As discussed in Section~\ref{sec:background}, both gem5 and SST are \emph{event-driven} simulators.
    We use the parallel event queue in SST as the main simulation event queue.
    On each cycle, SST calls into all of the gem5 simulator instances to advance their simulation.
    The parallel event simulation of SST guarantees that all of the gem5 simulations are constantly in sync with one another.
    Since SST is natively parallel, we should be able to scale the simulation to many nodes and fully use all of the cores on the host system.

    
    The rest of the section presents our approach to simulating \texttt{CXL.mem} with system-level fidelity across multiple hosts.
    Further, we describe the software extensions to gem5 and guide the user on how to get started with the interface.
    

    \subsection{Implementation}
        We organize the implementation discussion around two challenges: \textit{(C1)} initialization and \textit{(C2)} fast-forwarding into the region of interest.

        The first challenge, \textit{i.e.}, initialization, is making disaggregated memory OS-visible in a robust, reproducible way.
        Based on existing research, we opt for exposing the device as a NUMA node when pooling remote memory~\cite{tpp,pond}.
        For sharing, however, we expose the remote memory as a direct access character device~\cite{famfs}.
        
        While SST provides parallelism to gem5 simulations, booting up an OS with \texttt{systemd} enabled in gem5 can still take hours~\cite{bruce2021enabling}.
        Further, in Section~\ref{sec:npb}, we show that contemporary HPC workloads could be up to 85 GiB of application memory, which can take even longer just to allocate the memory.
        Therefore, we need to incorporate gem5's functional fast-forwarding and checkpointing with SST to jump to the ROI for feasibility.

        \begin{figure}[!b]
            \begin{center}
                \mbox{
                    \subfigure[CPU-less NUMA node via \texttt{numactl --hardware} inside gem5. The backend memory is simulated in SST.]
                        {
                        \label{fig:numa-cpuless}
                        \includegraphics[width=0.48\columnwidth]{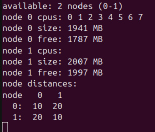}
                        }
                    \hspace{0.5ex}
                    \subfigure[Initialization of a DAX character device in the guest OS inside full-system gem5. The backend memory is simulated in SST.]
                        {
                        \label{fig:dax-device}
                        \includegraphics[width=0.4\columnwidth]{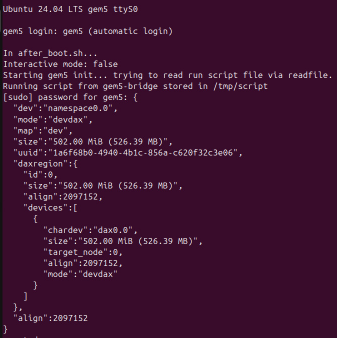}
                        }
                }
                \caption{System-level view of the remote memory ranges.}
                \label{fig:set-system}
            \end{center}
        \end{figure}


        \subsubsection{Challenge 1: Initialization}
            The first challenge in implementing memory disaggregation is the initialization and mapping of memory ranges across multiple independent system nodes.
            The CXL 3.1 specification addresses this by exposing the fabric-attached memory to the OS as distinct NUMA nodes~\cite{cxl-3.1-spec}.
            During initialization, system nodes can either allocate memory slices at boot time or via runtime hot-plugging. 
            Regardless of the mechanism, the OS remains agnostic of the underlying hardware topology, interacting with the disaggregated pool through standard NUMA-aware memory management.

            \begin{figure}[t]
                \centering
                \includesvg[width=0.96\columnwidth]{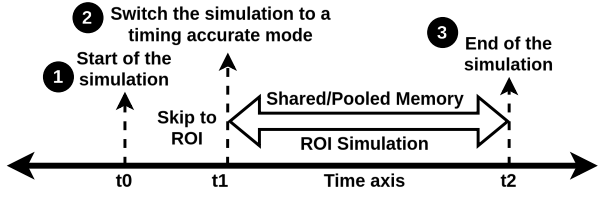}
                \caption{Ideal simulation timeline.}
                \label{fig:ideal-sim}
            \end{figure}

            \begin{figure}[!b]
                \centering
                \includesvg[width=0.96\columnwidth]{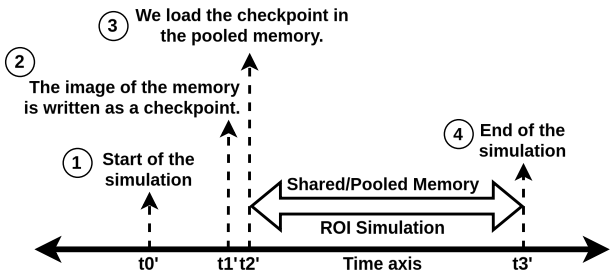}
                \caption{The simulation timeline for \papertitle. We fast-forward the simulation in gem5 to the ROI, take a checkpoint and restore the simulation in gem5 and SST using a timing-based model.}
                \label{fig:our-sim}
            \end{figure}
            While the CXL specification distinguishes between Host Physical Addresses (HPA) and Device Physical Addresses (DPA), our current implementation utilizes a simplified identity mapping with fixed address ranges.
            These ranges are hardcoded for each system node.

            In \papertitle, the gem5 simulator executes an unmodified Linux kernel.
            Consequently, the simulator must provide accurate address details via emulated hardware structures to allow the OS to establish valid software address ranges.
            On an x86 system, we define separate memory ranges using E820 entries.
            NUMA topology on x86 is defined via the Static Resource Affinity Table (SRAT). 
            Whereas, on an ARM/RISC-V system, memory ranges and NUMA affinity are defined via the Device Tree Blob (DTB).
            Figure~\ref{fig:numa-cpuless} shows how CPU-less NUMA memory node appear in full-system gem5.
            In both cases, we statically partition memory based on the specific requirements of the target workload.

            A critical distinction exists between pooling and sharing in our framework.
            To facilitate sharing, the remote memory must be exposed as a character device to prevent the kernel from zeroing out the memory and destroying shared data.
            Following the approach of FAMFS~\cite{famfs}, we map the remote memory as a DAX region (\texttt{/dev/dax}).
            This allows user processes to bypass the page cache and access the pool via direct, byte-addressable mappings.
            Figure~\ref{fig:dax-device} shows how DAX devices are initialized in full-system gem5.
            In \papertitle, standard NUMA policies are reserved strictly for pooling and are not utilized for shared memory regions.

        \subsubsection{Challenge 2: Simulating the Region of Interest}

            The integration of gem5 and SST presents a significant synchronization challenge.
            While SST excels at parallelizing multiple gem5 instances across MPI ranks, it lacks a shared backing store for simulated memory.
            Because system nodes may reside on different ranks, direct functional access, which is common in gem5 for fast-forwarding or emulating non-essential devices, becomes computationally expensive as it requires scheduling global SST events for every memory transaction.

            Ideally, a disaggregated simulation should transition seamlessly from functional initialization at time $t0$ (Action~\encircle{1}) to timing-accurate execution across all nodes (Figure~\ref{fig:ideal-sim}).
            This requires a global synchronization point at $t1$ (shown as Action~\encircle{2}) to switch from functional to timing models while maintaining a consistent view of the shared memory resource.
            Action~\encircle{3} represents business as usual as we proceed to collect and analyze the statistics of the simulation.



            The lack of a shared functional backing store in a distributed SST environment makes this ``ideal'' transition difficult to implement directly.
            To overcome this, \papertitle employs a two-phase simulation strategy using checkpointing (Figure~\ref{fig:our-sim}).

            We decouple the simulation into an initialization phase and a timing-accurate ROI (Region of Interest) phase:
            \begin{enumerate}
                \item \textbf{Functional Initialization (gem5-only):} At time $t0'$, we start each system node in gem5 using either the \texttt{AtomicSimpleCPU} or \texttt{KVM} CPU for high-speed fast-forwarding.
                During this phase, the memory is modeled functionally.
                Once the simulation reaches the ROI, we capture a memory checkpoint (Action~\wcircle{2}). This process is repeated for each of the $N$ nodes to establish their unique memory ranges.
                \item \textbf{Timing-Accurate Co-Simulation (gem5 + SST):} We restore the $N$ system node checkpoints within the integrated gem5-SST environment.
                At this stage ($t2'$), we switch to timing-accurate O3 CPU and memory models.
                Action~\wcircle{2} and Action~\wcircle{3} together form the equivalent of Action~\encircle{2}.
                This restoration acts as the global synchronization point, allowing all nodes to proceed in parallel until the simulation concludes at $t3'$ (Action~\wcircle{4}).
            \end{enumerate}

            This checkpoint-based approach allows \papertitle~to leverage the fast-forwarding capabilities of gem5 while maintaining the parallel scalability and timing accuracy provided by SST.

        \subsection{Software Details of \papertitle}
            \texttt{ExternalMemory} SimObject is based on the \texttt{OutgoingRequestBridge}, which, instead of exposing SST as an I/O range, exposes it as an \texttt{AbstractMemory} SimObject.
            This shared memory range is simulated using the \texttt{memHierarchy} element in SST, which has a \texttt{malloc}-based backing storage.
            
            \subsubsection*{Using \papertitle}
                \label{subsec:howto}
                The source code of \papertitle~is upstreamed to 
                \href{https://github.com/darchr/cxl-clustersim}{github}\footnote{\href{https://github.com/darchr/cxl-clustersim}{https://github.com/darchr/cxl-clustersim}}.
                We extend gem5's standard library (\texttt{stdlib}) to incorporate our proposed changes.
                gem5's \texttt{stdlib} provides users with pre-configured motherboard-like constructs that can be interfaced with plug-and-play components, including but not limited to different cache hierarchies, memory types, \textit{etc.}
                We extend the \texttt{ArmBoard}, \texttt{RiscvBoard}, and, the \texttt{X86Board} to implement memory disaggregation.
                The new boards are called \texttt{ArmComposableMemoryBoard}, \texttt{RiscvComposableMemoryBoard}, and, \texttt{X86ComposableMemoryBoard}.
                Instead of having one single memory range, we have two.
                The second memory range corresponds to an \texttt{ExternalRemoteMemory stdlib} component, which encapsulates \texttt{ExternalMemory}.
                This allows a seamless integration of gem5 and SST.
                Runscripts running with a detailed timing-based CPU automatically switch to the gem5 and SST configuration, where memory addresses bound to the remote node are serviced from SST.
            
\begin{lstlisting}[language=python, caption=gem5 runscript for taking checkpoints,label=code:take-ckpt]
# Model the local memory in gem5
local_memory = DualChannelDDR4_2400(size="4GiB")
# Map 4 GiB to 6 GiB to the remote node
remote_memory = ExternalRemoteMemory(
addr_range="4294967296,6442450944",
use_sst_sim=False)
# Use a KVM-based CPU to fast-forward the sim.
processor = SimpleProcessor(
cpu_type=CPUTypes.KVM,
isa=ISA.ARM, num_cores=4)
\end{lstlisting}

\begin{lstlisting}[language=python, caption=Restoring checkpoints on an SST-side script. The user can change the \texttt{disaggregated\_memory\_latency} (\textit{xx} nanoseconds (\textit{xxns})),label=code:restore-ckpt]
# add latency to memory requests going to SST.
disaggregated_memory_latency = "xxns"
# Specify to use SST
use_sst_sim = True
cpu_type = "o3"         # CPUTypes.O3
gem5_run_script = "../../disaggregated_memory/configs/arm-main.py"
\end{lstlisting}

                Code Listing~\ref{code:take-ckpt} shows a sample code to take a checkpoint in \papertitle.
                This represents a system with a slice of remote memory allocated from 4 GiB to 6 GiB in the address map.
                When restoring these checkpoints in SST, the SST-side script needs to be configured accordingly.
                This is shown in Code Listing~\ref{code:restore-ckpt}.
                In the current iteration of the code, we have automated the entire simulation end-to-end.

%% file: Text/eval_v2.tex
\section{Evaluation}
    \label{sec:eval}
    This section describes the evaluation platform used for the experiments.
    As mentioned before, \papertitle~uses gem5~\cite{gem5} and SST~\cite{sst}.
    The system nodes are modeled in gem5.
    Each of these is a comprehensive system, equipped with its own CPUs, caches, and dedicated local memory.
    They have their own independent operating system (OS).
    The parameters of the system node that we modeled are given in Table~\ref{tab:params}.



    \begin{table}[h]
        \caption{Simulation Parameters}
        \label{tab:params}
        \centering
        \begin{tabular}{cc}
            \hline
            \textbf{Parameter} & \textbf{Specification} \\ \hline \hline
            CPU type & Out-of-Order \\
            ISA & ARM (unless specified) \\
            CPU core count & 8 \\
            Frequency & 4 GHz \\ \hline
            L1 (I/D) cache per core & 32 KiB / 32 KiB \\
            L2 cache per core & 512 KiB \\
            L3 cache & 8 MiB \\ \hline
            Memory technology & DDR4 DRAM \\
            Memory frequency & 2400 MHz \\
            Number of local channels & 1 \\
            Number of remote channels & 4 \\
            Local memory size & 8 GB \\
            Remote memory size & variable/node \\ \hline
            Operating system & Ubuntu 22.04.4 \\
            Kernel version & 5.4.49 \\ \hline
            \end{tabular}%
    \end{table}

    \subsection{Calibration}
        \label{subsec:calibration}
        We validate and calibrate the basic model of the remote memory node using synthetic traces generated at the system node.
        The objective here is to calibrate our remote memory to a real-world memory device.
        Our objective is to model a 2400 MHz 4-channel memory with a peak theoretical bandwidth of 76.8 GB/s.
        The system that we used for the same has a synthetic traffic generator connected directly to a cross-bar switch, connected to the remote memory.
        We experiment by sending linearly generated synthetic read traffic using gem5's traffic generators.
        The memory is modeled in SST.
        This allowed us to find the maximum sustained and practical bandwidth without causing bank conflicts.
        The average total bandwidth reported was 59.6 GB/s.
        This is 77.5\% of the peak theoretical bandwidth of the DRAM device that we are trying to model.
        We consider this value the baseline bandwidth of the remote memory.

    \input{Text/stream}
    \input{Text/npb}
    \input{Text/gapbs}

%% file: Text/stream.tex
\subsection{Case Study I: Basic Full-System Experiments}

    We use the popular benchmark STREAM~\cite{stream-article, stream-techrep} to validate the full-system capabilities of our platform.
    STREAM is a synthetic benchmark program that measures the total memory bandwidth.
    There are four kernels in STREAM: copy, scale, add, and triad.
    The first two kernels work with two arrays, and the latter two kernels use three arrays.
    We set the size to 64 MiB for each of these arrays in our experiments. 
    
    For the full-system experiments, we use the Linux utility \texttt{numactl} to pin the program's memory only to the local memory (local), interleaving the pages between the two memory devices (interleave), or pinning all data to the remote memory (remote).
    The local memory has a single DDR4 DRAM channel with a peak theoretical bandwidth of 19.2 GB/s.
    Each system node assigned a 1 GiB range of memory for this experiment.

    We also added gem5's magic instruction-based annotations for the region of interest (ROI) at the beginning and end of each of the four kernels.
    The changes we made to the program allow us to take and restore checkpoints in gem5 and measure the bandwidth of each kernel individually.


    \subsubsection{Validating \papertitle}

        \begin{figure}[t]
            \begin{center}
                \mbox{
                    \subfigure[Baseline and observed bandwidth reported by the STREAM kernels at the external memory per kernel.]
                        {
                        \label{fig:theoretical-bw-comparison}
                        \includegraphics[width=0.22\textwidth]{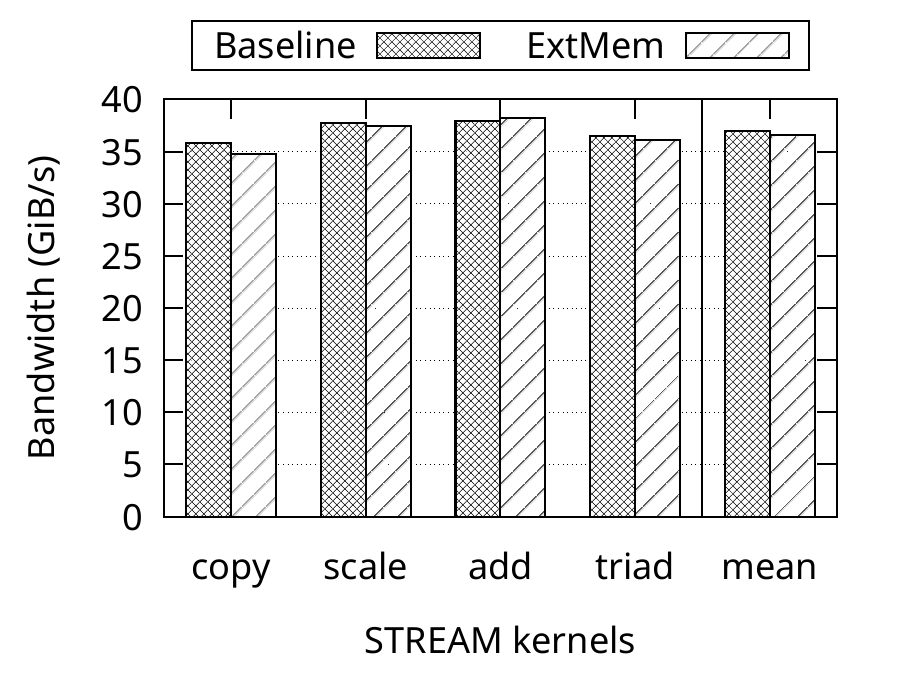}
                        }
                    \hspace{0.5ex}
                    \subfigure[Comparing the total baseline bandwidth seen in Section~\ref{subsec:calibration}, at the external memory, and at the remote memory controller.]
                        {
                        \label{fig:stream-at-three-places}
                        \includegraphics[width=0.22\textwidth]{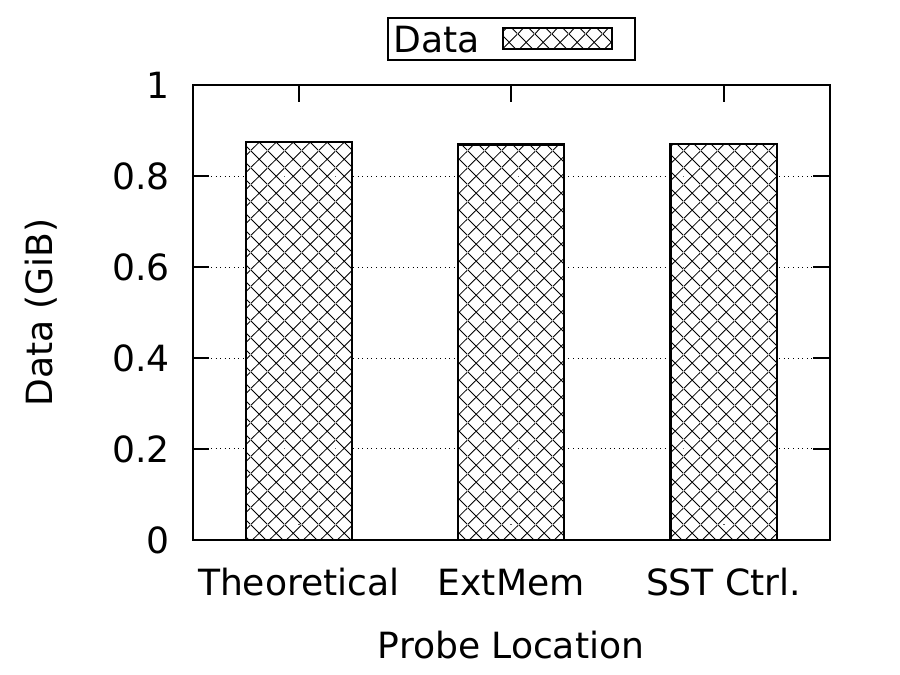}
                        }
                }
                \caption{Statistics used to validate the correctness of \papertitle. We show the baseline bandwidth of the system and compare it against the observed bandwidth of a single system node cluster.}
                \label{fig:set-1}
            \end{center}
        \end{figure}

        \begin{figure*}[t]
            \begin{center}
                \mbox{
                    \subfigure[When \texttt{numactl} pinned the program to the local memory.]
                        {
                        \label{fig:numactl-local}
                        \includegraphics[width=0.22\textwidth]{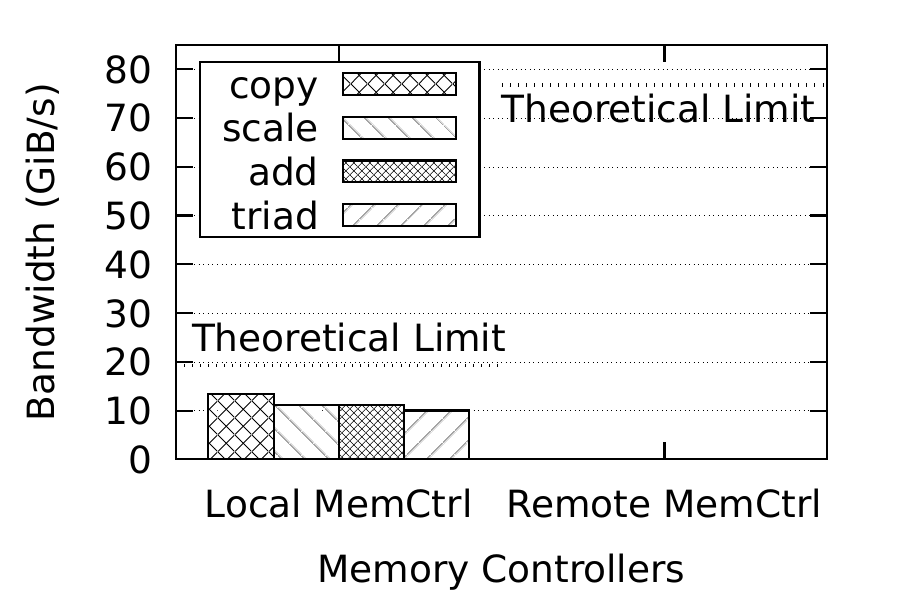}
                        }
                    \hspace{0.05ex}
                    \subfigure[When \texttt{numactl} interleaved the program to use both the local and the remote memory.]
                        {
                        \label{fig:numactl-interleaved}
                        \includegraphics[width=0.22\textwidth]{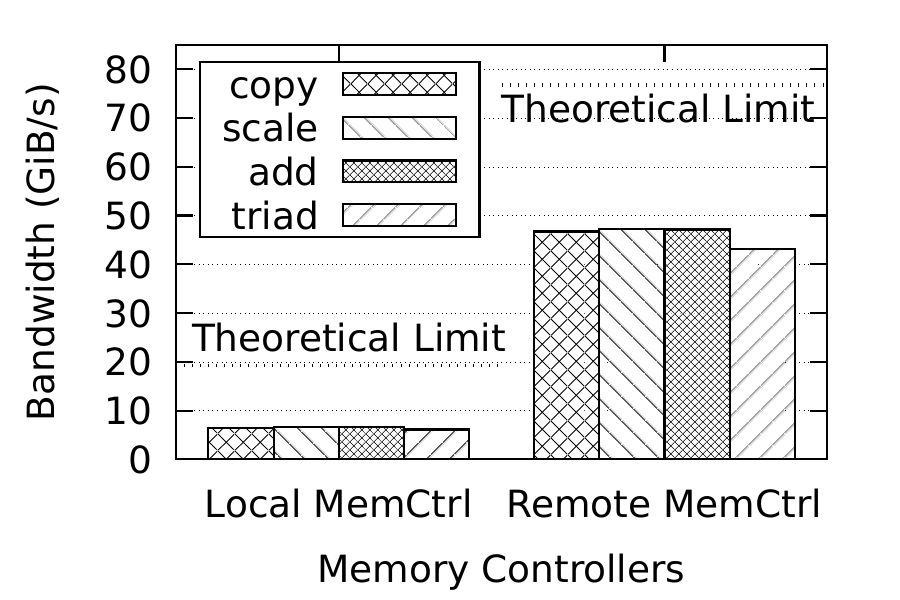}
                        }
                    \hspace{0.05ex}
                    \subfigure[When \texttt{numactl} pinned the program to the remote memory.]
                        {
                        \label{fig:numactl-remote}
                        \includegraphics[width=0.22\textwidth]{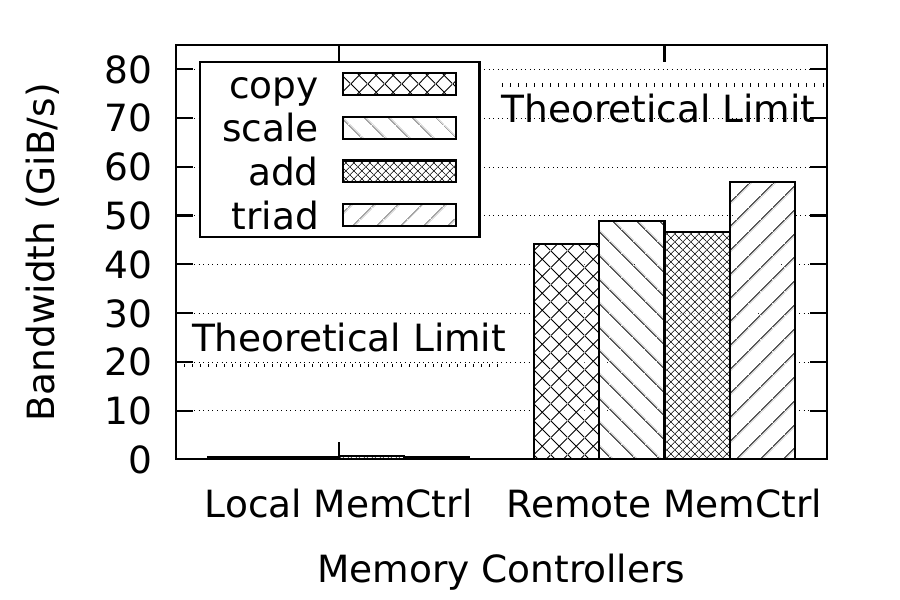}
                        }
                    \hspace{0.05ex}
                    \subfigure[Reported average bandwidth per kernel at each of the memory controllers. The program was interleaved between the two memories.]
                        {
                        \label{fig:numactl-interleaved-2}
                        \includegraphics[width=0.22\textwidth]{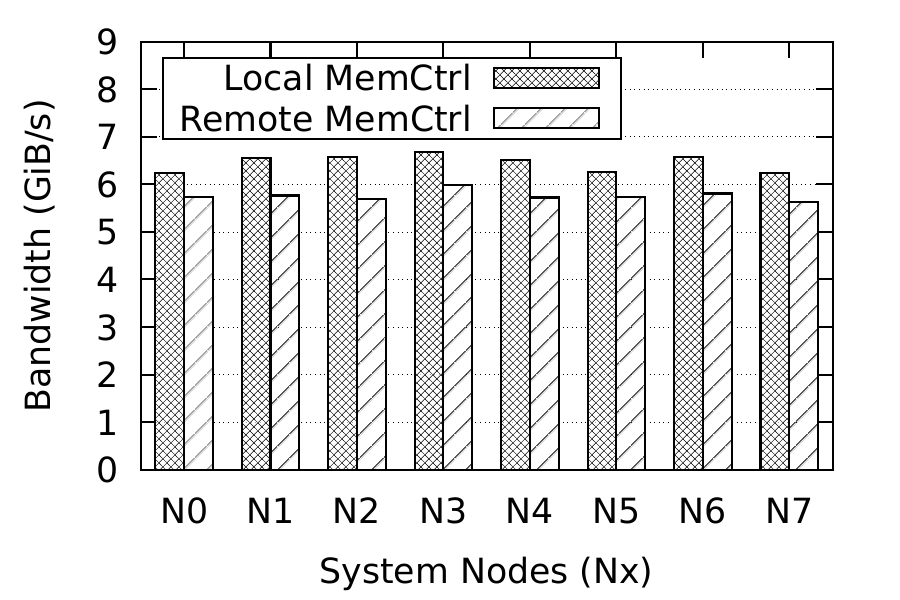}
                        }
                }
                \caption{The total bandwidth was measured at each of the memory controllers.
                There were 8 system nodes in the cluster running STREAM.}
                \label{fig:set-2}
            \end{center}
        \end{figure*}
        The first set of results are shown with a single system node connected to a remote node (Figure~\ref{fig:set-1}).
        In Figure~\ref{fig:theoretical-bw-comparison}, we show the baseline and reported bandwidth of the STREAM kernels when \texttt{numactl} pins the kernels to the remote memory.
        The \texttt{ExternalMemory} (i.e., the simulated CXL link) reports how much data was moved over this link.
        We expect to see the baseline and the reported numbers to be \textit{similar} and not exact.
        This is because there are factors such as caching and prefetching affecting the simulation.
        The difference reported is less than 1.0\%.
        This suggests that \papertitle~is set up correctly from gem5's side.
        Further, Figure~\ref{fig:stream-at-three-places} shows the total bandwidth reported by the SST's memory controller.
        SST reports the bandwidth off by 0.1\%, which is attributed to resetting the statistics using the gem5 annotations.

        \begin{framed}
            \textbf{Observation 1}: The baseline and the reported bandwidth using \papertitle~are within 0.99\%. This suggests that our memory disaggregation model is correct.
        \end{framed}

    \subsubsection{Running STREAM across multiple system nodes}
        \label{subsec:stream-across-nodes}

        The bandwidth reported at each of the memory controllers in an eight-node cluster is shown in Figure~\ref{fig:set-2}.
        We show the ``Local MemCtrl'' bandwidth as the average bandwidth of each of the system nodes, and the ``Remote MemCtrl'' bandwidth as the bandwidth at the remote memory system.
        We used \texttt{numactl} to pin STREAM to use the local memory (Figure~\ref{fig:numactl-local}), interleave between the local and the remote memory (Figure~\ref{fig:numactl-interleaved}), and pin the application memory to the remote memory (Figure~\ref{fig:numactl-remote}).
        We do not see any significant traffic at the remote link when we pin the process to the local memory.
        Likewise, the traffic is exclusive to the remote memory node when the process is pinned to the same remote memory.

        In the interleaved memory setup, application bandwidth is heavily constrained by the slower remote memory node.
        Figure~\ref{fig:numactl-interleaved-2} shows that the effective memory bandwidth per system node peaks at only 6.45 GiB/s, a significant drop from the 11.4 GiB/s baseline of strictly local memory (Figure~\ref{fig:numactl-local}).
        Because interleaved NUMA distributes memory allocations across both local and remote nodes, the performance is bottlenecked by the shared remote link. 
        Specifically, the remote node delivers a measured total bandwidth of 46.04 GiB/s across all eight compute nodes (out of the 59.6 GiB/s, as discussed in Section~\ref{subsec:calibration}). 
        This equates to an average remote bandwidth of just 5.75 GiB/s per node, which fundamentally throttles the overall memory throughput of the interleaved execution.


        \begin{framed}
            \textbf{Observation 2}: The \texttt{numactl} can be effectively used in \papertitle. The remote node behaves close to a memory blade, where it is constrained by its own specifications.
        \end{framed}


    \subsubsection{CXL Latency}

\begin{figure}[!b]
            \centering
                \includegraphics[width=0.46\textwidth]{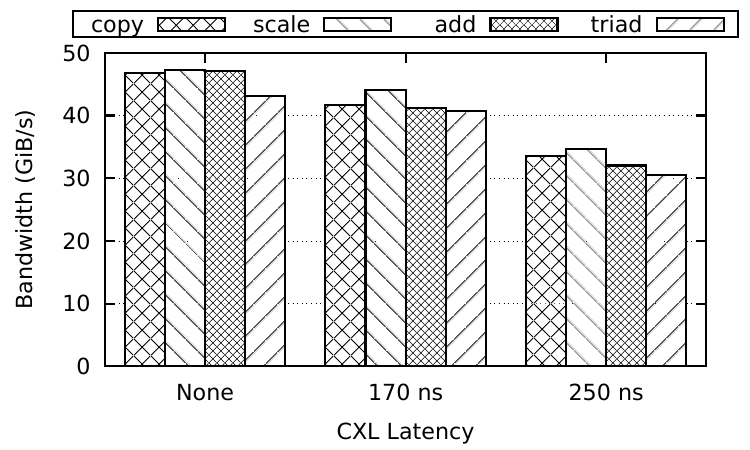}
                \caption{The bandwidth reported by each of the STREAM kernels when the CXL latency was varied. There were 4 system nodes.}
                \label{fig:cxl-latency}
        \end{figure}

        Our next experiment adds latency to the link to the remote memory system (i.e., to model CXL latency).
        In this experiment, we model four system nodes cluster instead of eight in the previous experiment.
        We use SST to model this CXL latency.
        Sharma \textit{et al}. reported that the initial CXL latency numbers reported were within the range of 170 ns to 250 ns~\cite{dassharma}.
        Therefore, we show the bandwidth of the system running STREAM remotely by varying the CXL latency in Figure~\ref{fig:cxl-latency}.

        In the experiment, we can clearly see that the total bandwidth is the highest when there is no CXL latency.
        The bandwidth decreases as we increase the latency.
        The impact on the bandwidth from no latency to 170 ns is 8.95\%, and to 250 ns is 29\%.
        Compared to the 170 ns case, the dip in performance for 250 ns is 22\%.
        Backpressure is implemented on the SST side for the external CXL links.

        \begin{framed}
            \textbf{Observation 3}: The disaggregated memory latency has an impact on the bandwidth of the remote node.
        \end{framed}

        \begin{figure}[!b]
            \begin{center}
                \mbox{
                    \subfigure[The amount of host memory required to simulate 1 to 16 system nodes. We report the maximum resident set size (Max. RSS) and approximate maximum global resident set size (Global RSS) of the simulation.]
                        {
                        \label{fig:rss}
                        \includegraphics[width=0.22\textwidth]{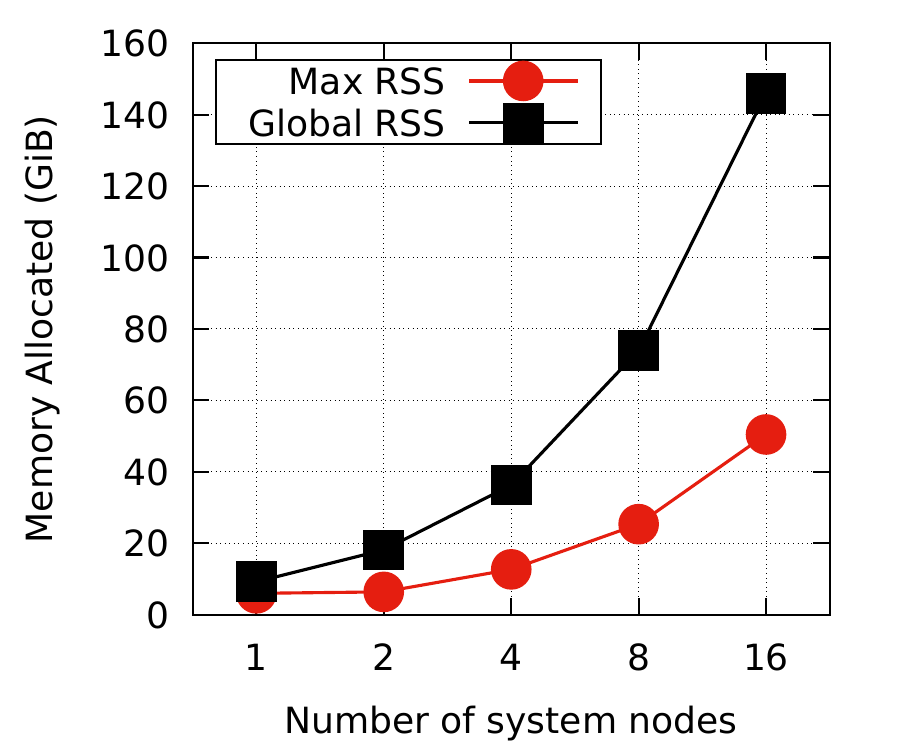}
                        }
                    \hspace{0.05ex}
                    \subfigure[Reported parallel efficiency of the cluster when scaling system nodes from 1 to 16 in powers of 2. There is one extra MPI rank for the remote memory node.]
                        {
                        \label{fig:wallclock}
                        \includegraphics[width=0.22\textwidth]{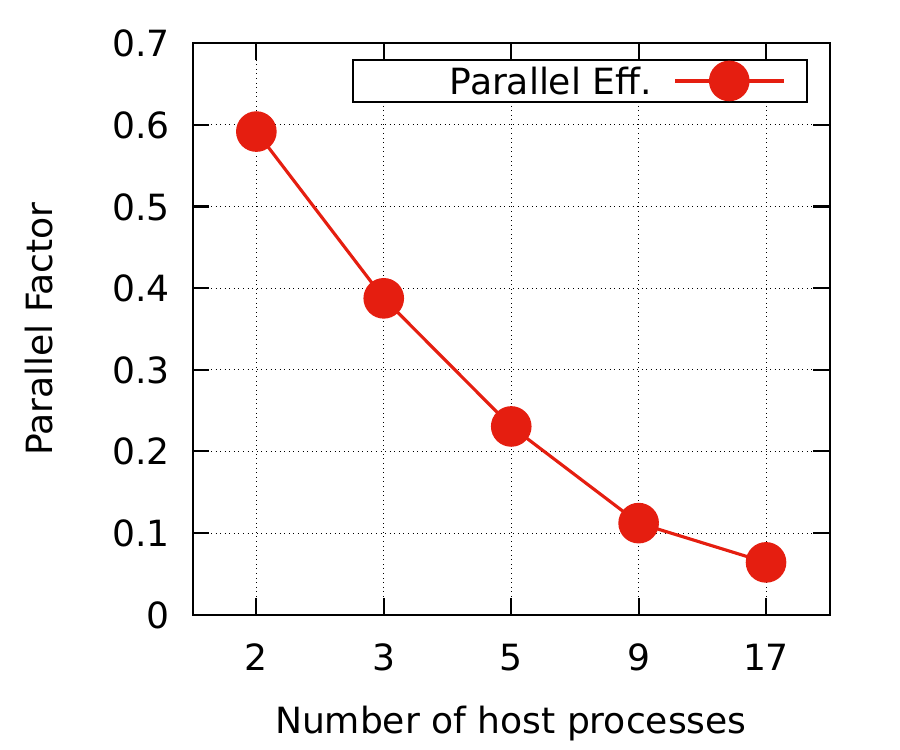}
                        }
                }
                \caption{Reported host system statistics when the number of system nodes varied.}
                \label{fig:set-4}
            \end{center}
        \end{figure}

    \subsubsection{Host System and Parallelization Statistics}
        The host system (i.e., the system that is executing the gem5 simulations) also constitutes an important part of the simulations.
        We used an ARM-based Neoverse-N1 architecture with 160 cores.
        The system was equipped with 512 GiB of total memory.
        In this section, we report the simulation statistics for the host system when we increased the number of system nodes from 1 to 16 in powers of 2.
        Collectively, these statistics are shown in Figure~\ref{fig:set-4}.

        The first statistic is the amount of host memory required.
        Figure~\ref{fig:rss} shows the memory requirements on the host system for simulating system nodes from 1 to 16.
        Since SST is an MPI-based software, we report the MPI statistics of SST at the end of each simulation.
        The memory required by one MPI process is given by the \textit{maximum resident set size}, or simply \textit{RSS}.
        The total amount of memory required for the entire simulation cluster is given by \textit{Approximate global RSS size}.
        Our simulations for 8 system nodes allocated 25.3 GiB of memory per system node.
        The global RSS size was reported to be 73.9 GiB.
        For 16 system nodes, the global RSS size was reported to up to 145.96 GiB.
        The trend is observed to be linear for both RSS and global RSS as we scale with the number of system nodes.

        The other statistic that is crucial for the \papertitle~is the wall clock or the simulation runtime.
        Since SST enables parallel simulation, we would hope that as we increased the number of simulated system nodes the performance of the overall simulation would not degrade.
        However, we found that as we increased the number of system node, the wall clock time increased significantly.

        Figure~\ref{fig:wallclock} shows the parallel efficiency when scaling system nodes.
        We use the single system node running STREAM in gem5 memory as our baseline case.

        \begin{figure}[t]
            \centering
                \includegraphics[width=0.48\textwidth]{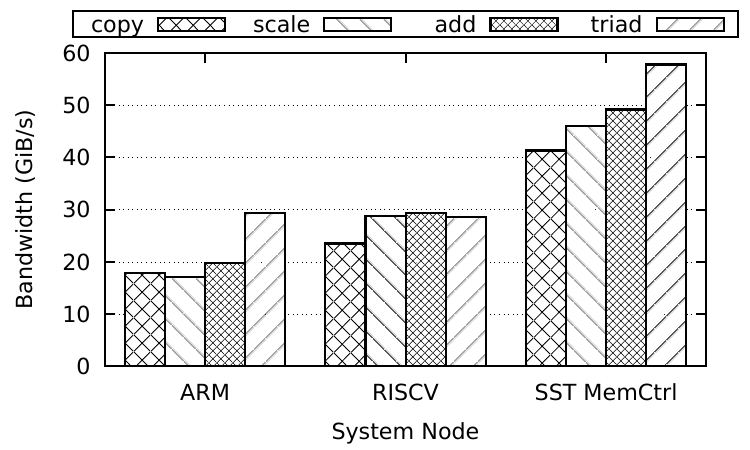}
                \caption{Bandwidth reported by the two system nodes and at the SST's memory controller when two heterogeneous system nodes ran STREAM remotely.}
                \label{fig:mix-isa}
        \end{figure}

        Parallel efficiency (PE) is defined in Equation~\ref{eq:pe}.
        Parallel efficiency is 1.0 if the time running $N$ system nodes on $N$ threads is the same as the serial execution of a single system node.
        Parallel efficiency below 1.0 implies synchronization or other overheads in the parallelization of the simulation.
        \begin{equation}
            \label{eq:pe}
            \texttt{PE} = \frac{1}{\texttt{NUM}_{\texttt{processes}}}\times \frac{\texttt{TIME}_{\texttt{gem5 only}}}{\texttt{TIME}_{\texttt{\papertitle}}}
        \end{equation}

        We start by showing the parallel efficiency with two host processes as the first point in Figure~\ref{fig:wallclock}.
        In this case, we have one host process executing the gem5 simulation (the system node) and one process executing the remote memory system.
        In this case, we see a 16\% improvement in wall clock time compared to running gem5 alone, which implies that there is some parallelism between the compute-based node and the memory node.
        However, as we continue to scale the system nodes, we see an degradation in the parallel efficiency.


        The PE of a two node cluster is 0.38.
        The number dips to 0.06 for the 16 system nodes case.
        The reason for this decrement is the bottleneck caused at the SST-core for the remote memory node.
        We have pinned the STREAM kernels to the remote memory only, which makes the simulator serialize the incoming memory requests across all the system nodes for the remote memory rank.

        Later in Figure~\ref{fig:gapbs-split}, we show that the memory footprint of these real-world workloads is distributed between local and remote memory.
        Because the workloads do not rely exclusively on the remote node, the volume of remote memory accesses is limited, which significantly reduces the overall simulation time.

        While the parallelism is not efficient, there is a slight speedup (about 1.09$\times$ speedup) over the serial version at 17 host processes.
        This provides an opportunity to improve \papertitle's efficiency in the future.

        \begin{framed}
            \textbf{Observation 4}: The parallel efficiency is better than linearly scaling the system nodes.
            There are ample opportunities to improve this further.
        \end{framed}

    \subsubsection{Mixing ISAs}
        The final set of experiments for STREAM shows how heterogeneous ISA system nodes can pool memory from the remote memory node.
        The memory device is agnostic of the ISA.
        As long as there are incoming memory requests within its range, the remote node is expected to function.
        A system node should only be aware of memory ranges.
        Therefore, we set up an interesting experiment where we use an ARM-based system node and a RISC-V-based system node.
        Except for the ISA, every other parameter of the systems is identical, defined in Table~\ref{tab:params}.

        We execute STREAM and pin it to the remote memory.
        Figure~\ref{fig:mix-isa} shows the reported bandwidth, and we find the results are similar to what we have already seen in Section~\ref{subsec:stream-across-nodes}.
        This adds a new angle for researchers to explore using \papertitle.
        This is especially interesting as the data for the RISC-V node shows that it was able to exploit the remote node's bandwidth by 31\% more than that of the ARM system node.
        This opens the door to studying and comparing how the differences in the core architecture can drive memory bandwidth utilization.

        \begin{framed}
            \textbf{Observation 5}: The remote node is agnostic of the system node's architecture. \papertitle, therefore, can be used to study heterogeneous CPUs alongside heterogeneous memories.
        \end{framed}

%% file: Text/npb.tex
\begin{table}[t]
    \centering
    \caption{NPB Workloads Working Set Sizes in class D}
    \begin{tabular}{cc}
    \hline
    Applications & Working Set Size (GiB) \\ \hline \hline
    bt          & 11    \\ 
    cg          & 17    \\ 
    ep          & 1     \\ 
    ft          & 85    \\ 
    mg          & 27    \\ 
    sp          & 12    \\ 
    ua          & 8     \\ \hline
    \end{tabular}
    \label{tab:npbWorkloadsSizes}
\end{table}

\subsection{Case Study II: Memory Pooling} 
\label{sec:npb}

In this study, we evaluate the performance of disaggregated systems while running the NAS Parallel Benchmarks (NPB)~\cite{npb}. 
The NPB is a set of scientific programs designed to help evaluate the performance of parallel supercomputers, and they are widely used to assess computational and communication performance in various scientific and engineering domains.

For this study, we use \papertitle~to configure two different systems to study the memory stranding issue in datacenters, for which memory disaggregation solution has been proposed and implemented.
To this end, we consider a cluster that consists of 7 system nodes (host in CXL spec), each with 8 cores, a cache hierarchy, and a connection to the remote memory node.
The system configuration parameters are described in Table~\ref{tab:params}.
We chose a subset of NPB workloads, each of them running on a separate system node.
Table~\ref{tab:npbWorkloadsSizes} shows the working set size of the used applications from NPB, rounded up to the nearest GiB.
We chose class D as the input size for the NPB workloads, as it is sufficiently large for the systems we configure.
We simulate these systems for 500~ms in full-system mode that captures the interactions of the entire system, including the operating system (TLBs, NUMA, huge pages, etc.) and the application behavior during runtime.

We define two different memory setups, as follows:
\begin{itemize}
  \item \textbf{No-NUMA}:
  In this configuration, each system node is equipped with 128 GiB of local memory.
  We choose 128~GiB based on the largest workload we test (ft, 85~GiB), rounded up to the nearest power of two.
  Given the ample local memory, all data required by the benchmarks fit entirely within the local memory.
  This setup removes the need for remote memory accesses, thereby reducing potential remote memory access latencies.
  However, it cannot guarantee that the system node will utilize the entire 128GiB of memory, as the working set size of the application may be smaller than the local memory size.

  \item \textbf{NUMA-Local-Preferred}:
  In this configuration, the system is equipped with a significantly smaller local memory of 8 GiB, supplemented by a shared pool memory of 160 GiB.
  This setup is designed to reflect a scenario where the local memory is insufficient to hold all the data, necessitating that a portion of the data be stored in remote memory.
  The system employs a NUMA policy with a preference for local memory access.
  Thus, it first starts to allocate memory from local memory.
  If the working set size of the application is larger than the local memory, the remaining portions of the workload inevitably reside in the remote memory, leading to increased latency and potential performance degradation.
\end{itemize}

\begin{figure}[t]
    \centering
    \includegraphics[width=0.96\columnwidth]{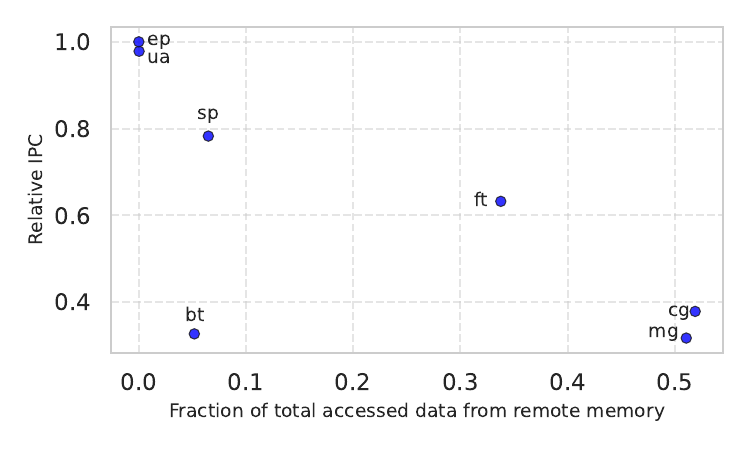}
    \caption{NPB workloads' relative IPC of NUMA-Local-Preferred memory setup compared to the No-NUMA, against the fraction of working set size that was accessed in the remote memory.
    Y-axis starts at 0.3.}
  \label{fig:npb-relIPC}
\end{figure}


Figure \ref{fig:npb-relIPC} shows the relative instruction per cycle (IPC) of the NPB workloads in the NUMA-Local-Preferred memory setup compared to the No-NUMA memory setup on the Y-axis.
It also shows the fraction of the working set size that was accessed in the remote memory on the X-axis.
The figure shows that the relative IPC of the workloads decreases as the fraction of the working set size accessed in the remote memory increases, due to the longer latency of accessing the remote memory node and potential interference once the application accesses the remote memory simultaneously.
$sp$, $ft$, $cg$, and $mg$ show a decrease in their relative IPC as the fraction of accessed remote data increases.
The larger the fraction of the working set size accessed in the remote memory, the higher the performance degradation.
For instance, 52\% of the data of $mg$ (working set size of 27~GiB) was accessed in the NUMA-local-Preferred setup was in the remote memory, which resulted in a 0.38 relative IPC, \textit{i.e.}, 62\% overall slowdown in its performance compared to No-NUMA setup.
Even though the relative IPC decreases, the amount of memory stranding at each system node's local memory is significantly reduced as well.
In the No-NUMA setup, each system node has 128~GiB of local memory, which is significantly higher than the  working set size of most of the workloads.
For example, in the $mg$ workload, the working set size is 27~GiB.
This means that 79\% of the stranded local memory is saved in the NUMA-Local-Preferred setup.

Figure \ref{fig:npb-relIPC} also shows that the relative IPC of the workloads is not significantly affected when the fraction of the working set size accessed in the remote memory is negligible.
$ep$ and $ua$ workloads do not show a significant decrease in relative IPC, as the fraction of accessed remote data for them is close to zero.
Note that the NUMA-Local-Preferred memory setup has 8~GiB of local memory.
As Table~\ref{tab:npbWorkloadsSizes} shows, the working set sizes of the $ep$ and $ua$ workloads are 1~GiB and 8~GiB, respectively, fitting entirely in the local memory.

Out of all the workloads, only $bt$ shows a decrease in relative IPC while the fraction of accessed remote data is not significant.
The anomalous IPC degradation of the \textit{bt} workload, despite a seemingly low fraction of remote data access, is driven by a combination of \texttt{numactl --preferred}, and its memory footprint and highly irregular access patterns.
\texttt{numactl --preferred} is a soft policy, which, unlike \texttt{numactl --mem-bind}, tries to use a preferred numa node, which may not yield positive results everytime.
For workloads with irregular access patterns like \textit{bt} and \textit{sp}, we saw varying results each time we ran the experiment.
On the other hand, workloads with regular access patterns or smaller remote memory footprint like \textit{mg}, \textit{ep} and \textit{ua} had the fraction of remote memory data similar across multiple runs.

\begin{framed}
  \textbf{Observation 6}: While remote memory access significantly degrades IPC due to increased latency and interference, the NUMA-Local-Preferred setup effectively minimizes memory stranding by rightsizing local allocations to actual workload demands.
\end{framed}

%% file: Text/gapbs.tex
\subsection{Case Study III: Memory Sharing}
    \label{sec:gapbs}
    To evaluate memory sharing, we study graph analytics workloads using GAPBS~\cite{gapbs}.
    CXL 3.0 enables memory sharing across hosts; in our framework, \papertitle~models this on x86 by mapping the remote memory node as a DAX region, akin to the FAMFS approach~\cite{famfs}, so user processes can access the remote memory via a direct, byte-addressable mapping without page cache mediation.
    This mirrors current programming practice with \texttt{/dev/dax} for heterogeneous memory systems and allows us to exercise cross-host sharing semantics in a controlled setting.

    \begin{figure}[t]
        \centering
        \includegraphics[width=0.96\columnwidth]{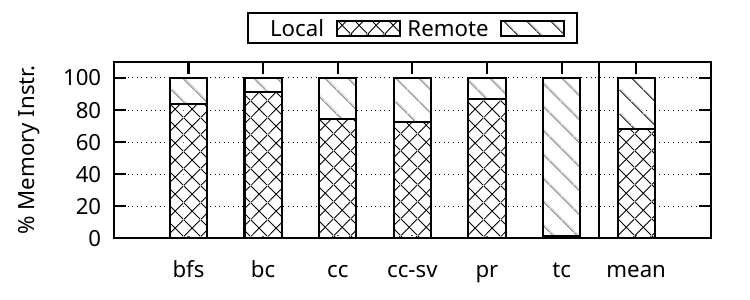}
        \caption{Share of retired memory instructions served by \textit{local} vs. \textit{remote} memory across all the GAPBS kernels under our sharing setup.
        Bars are normalized to 100\% per kernel.
        \textit{Remote} denotes requests serviced by the SST-modeled remote memory node, and \textit{Local} denotes the node's attached DRAM.
        The \textit{mean} bar reports the arithmetic average across kernels.
        }
    \label{fig:gapbs-split}
    \end{figure}

    We use a modified version of GAPBS
    ~\footnote{\href{https://github.com/darchr/shared-gapbs}{https://github.com/darchr/shared-gapbs}}
    that supports loading a single graph image into the remote memory node once and then mapping that image into multiple hosts.
    One designated ``\textit{allocator}'' host performs the initial allocation and population of the graph in the shared remote memory (the single writer), while the remaining hosts map the graph as read-only (multiple readers).
    This reflects a common analytics pattern where many readers traverse a static graph snapshot.
    We concurrently run six GAPBS kernels (\textit{bfs, bc, cc, cc\_sv, pr,} and \textit{tc}) each on a separate system node, all reading the same remote graph.
    We validated the correctness of the changes to the version of GAPBS by observing the output of each of the kernel with the unmodified version.

    Figure~\ref{fig:gapbs-split} breaks down the fraction of retired memory instructions served by local DRAM versus the shared remote blade across all six GAPBS kernels under our single-writer, multiple-reader  configuration.
    The split is kernel-dependent, confirming that a non-trivial portion of accesses are satisfied by the remote blade when the graph resides in shared memory, while private/stack activity remains predominantly local. 
    On average, 31.8\% of the total memory instructions are served from the shared remote memory node.

    We compare against a baseline configuration that runs each kernel on a single gem5-only system with private memory (no remote access), reporting instructions per cycle (IPC).
    In the shared-memory configuration, all kernels run in parallel across hosts, accessing the single, DAX-mapped graph in the remote memory node.
    We add remote memory path latency in SST (250 ns).
    Unless otherwise noted, per-kernel IPC is collected over each kernel's region of interest, and we also report a geometric mean across kernels for compactness.

\begin{figure}[t]
        \centering
        \includegraphics[width=0.96\columnwidth]{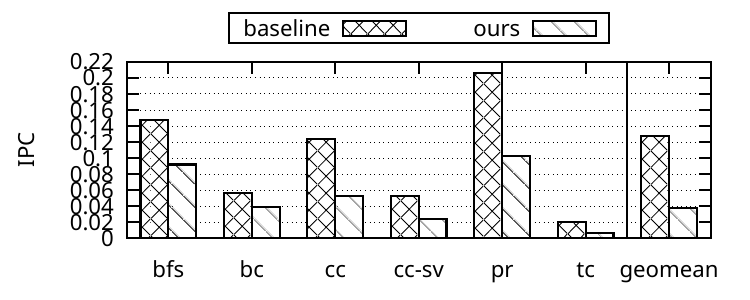}
        \caption{Reported IPC of GAPBS kernels in \papertitle, when the same unweighted synthetic graph was shared across all the kernels. We compare the IPC of \papertitle's experiments with a single system running one kernel without additional latencies.}
    \label{fig:gapbs}
    \end{figure}

    Figure~\ref{fig:gapbs} shows the results.
    As expected, moving the graph to the shared remote memory node reduces IPC relative to the baseline due to \textit{(1)} extra access latency on the CXL path and \textit{(2)} cross-host contention at the remote memory controller when kernels co-execute.
    Increasing the injected CXL latency will further degrade performance.
    Kernels with more irregular, pointer-chasing access patterns (bfs, pr) tend to be more sensitive than those with more streaming-friendly behavior, while compute-heavy kernels exhibit milder slowdowns.
    The geometric mean reflects these trends across kernels.

    This experiment demonstrates that \papertitle~can model cross-host sharing from user space via DAX-style mappings and capture interference effects in the shared remote memory node.
    Our current sharing model intentionally uses a single-writer/multiple-reader discipline with read-only mappings on readers; we do not model host-to-host cache-line coherence and keep it as future work.
    Additionally, the present CXL path omits link-level backpressure and switch modeling, which likely understates tail-latency growth and interference under heavy co-tenancy; we leave detailed switch/fabric and credit-based flow-control modeling to future work.

    \begin{framed}
        \textbf{Observation 7}: \papertitle demonstrates that CXL-based memory sharing via DAX mappings enables multi-host analytics at the cost of performance degradation driven by remote access latency and controller contention, particularly in pointer-chasing workloads.
      \end{framed}




%% file: Text/related.tex
\section{Related Work}


\label{sec:related}    
\textbf{gem5-based Distributed System Simulation.}
gem5 is a cycle-level simulator offering a rich amount of micro-architecture details while providing full-system simulation capability.
Bringing up multiple system nodes within a single gem5 simulation is a possibility, however only a single CPU thread is used for simulating all the gem5 nodes.
As a result, gem5-based distributed system simulation has been the target of many prior works extending the parallelization capability of gem5.

Alian \textit{et al.} propose dist-gem5~\cite{mohammad2017dist}, a modified version of gem5 supporting distributed system simulation via MPI.
The work allows multiple gem5 instances to run simultaneously, and facilitates the communication between the instances using an ethernet inter-node switch model.
Compared to our work, dist-gem5 requires maintaining the MPI-related handling, while our gem5-SST implementation offloads this responsibility entirely to SST.
COSSIM~\cite{tampouratzis2020novel} is another distributed gem5-based framework that is capable of simulating multiple system nodes, and the inter-node communication is primarily ethernet-based.
While our work also targets simulating distributed system, the communication of a disaggregated memory system is not ethernet-based.

Puri \textit{et al.} proposed DRackSim~\cite{dracksim}, where hosts are modeled in gem5, and the remote memory is modeled in DRAMSim2~\cite{dramsim2}. 
The framework relies on gem5's Syscall Emulation (SE) mode, which does not take system-level overheads or kernel-level memory management behaviors.

Recent frameworks like Cohet~\cite{cohet,cxl-dmsim} and Xerxes~\cite{xerxes} offer valuable micro-architectural models for single-node CPU-XPU coherence and CXL 3.0+ fabric routing in gem5, respectively.
gem5-CXL~\cite{gem5-cxl} on the other hand, focuses on optimizing localized memory parallelism.
Crucially, \papertitle is complementary to these efforts as we focus on multi-host simulation.
Both of these works' localized device pipelines and link-layer optimizations can be modularly plugged into our core gem5 component, allowing SST to seamlessly scale their innovations to explore cluster-wide memory pooling and multi-host resource contention.

SST is a discrete event-based simulation toolkit which comes with a MPI-based framework for parallelizing simulations and facilitating communication between simulations~\cite{sst}.
SST naturally complements the lack of parallelization in gem5, as well as offers the separation of the responsibility of handling MPI events to SST.
Unsurprisingly, there is a history of integration between gem5 and SST.
Hsieh \textit{et al.}~\cite{gem5sst} show an integration of gem5 and SST in 2012, demonstrating the scaling capability of scaling the number of system nodes using SST.
However, we are not aware of the status of the 2012 integration with more recent versions of gem5 and SST.
A more recent integration of gem5 and SST in 2021~\cite{sst2021} addresses the maintainability of gem5/SST integration by making the integration part of gem5 tests.
The work is demonstrated to be able to bring up multiple gem5 \emph{cores} running in parallel using a recent version of SST.
This prior gem5-SST combination was focused on providing the detailed core model of gem5 as a component to SST-based simulations without fast-forwarding or checkpointing.

\textbf{Disaggregated Memory System Performance Profiling.}
Sun \textit{et al.} have characterized the performance of CXL 1.1 devices~\cite{demystifying-paper}.
They highlight the difference between CXL 1.1 devices and the NUMA-based emulations.
The work is limited to understanding the performance characteristics of the CXL 1.1 protocol and related devices.
Our work captures the performance characteristics by modeling essential characteristics of CXL devices, such as CXL latency, while keeping the implementation simple to enable disaggregated memory system exploration.

%% file: Text/conclusion.tex
\section{Conclusion}
    \label{sec:conclusion}

    In this paper, we present \papertitle, a full system modeling and simulation infrastructure for performance evaluation and design space exploration of CXL-based disaggregated memory. 
    The proposed system has some limitations.
    Dynamic memory management via memory hot-plugging, and the details of CXL's coherence specification are not incorporated yet.
    This, along with modeling the CXL switch constitutes our plans for future work.
    One can take advantage of the framework infrastructures and extend to evaluate the CXL.io/CXL.cache-based systems.
    Even with these limitations, we expect \papertitle~to be a valuable asset to the computer architecture research community to benchmark and evaluate hardware/software co-design and optimization ideas in this rapidly evolving research topic.

%% file: Text/ack.tex
\section*{Acknowledgments}

We thank Ayaz Akram, Bobby Bruce, Mahyar Samani, William Shaddix, and the other members of DArchR who provided feedback on earlier versions of this work and contributed to the development of the gem5-SST integration.
This work was supported in part by the National Science Foundation under grant numbers 1925724, 2144883 and 2311888, Department of Energy under grant number DE-SC0021149, and the Laboratory for Physical Sciences.

LLMs were used for editorial purposes in this manuscript, and all outputs were inspected by the authors to ensure accuracy and originality.